\documentclass[superscriptaddress,twocolumn,prd,preprintnumbers,amsmath,amssymb,nofootinbib]{revtex4}

\usepackage{graphicx}
\usepackage{dcolumn}
\usepackage{bm}
\usepackage{epsfig}
\usepackage{epstopdf}
\usepackage{amssymb}
\usepackage{amsmath}
\usepackage{subfigure}
\usepackage{color}

\newcommand{\beq}{\begin{equation}}
\newcommand{\eeq}{\end{equation}}

\newcommand{\eq}{\mathrm{eq}}
\newcommand{\RH}{\mathrm{RH}}
\newcommand{\F}{\mathrm{F}}
\newcommand{\mpl}{m_{\mathrm{pl}}}
\newcommand{\sigunits}{\,\mathrm{cm^3\,s^{-1}}}

\definecolor{orange}{rgb}{1,0.5,0}

\begin{document}

\title{New Constraints on Dark Matter Production during Kination}

\author{Kayla Redmond}
\email{kayla.jaye@unc.edu}
\affiliation{Department of Physics and Astronomy, University of North Carolina at Chapel Hill, Phillips Hall CB3255, Chapel Hill, NC 27599 USA}
\author{Adrienne L. Erickcek}
\email{erickcek@physics.unc.edu}

\begin{abstract}
Our ignorance of the period between the end of inflation and the beginning of Big Bang Nucleosynthesis limits our understanding of the origins and evolution of dark matter. One possibility is that the Universe's energy density was dominated by a fast-rolling scalar field while the radiation bath was hot enough to thermally produce dark matter.  We investigate the evolution of the dark matter density and derive analytic expressions for the dark matter relic abundance generated during such a period of kination.  Kination scenarios in which dark matter does not reach thermal equilibrium require $\langle \sigma v \rangle < 2.7\times 10^{-38} \sigunits$ to generate the observed dark matter density while allowing the Universe to become radiation dominated by a temperature of $3 \, \mathrm{MeV}$.  Kination scenarios in which dark matter does reach thermal equilibrium require $\langle \sigma v \rangle > 3\times 10^{-26} \sigunits$ in order to generate the observed dark matter abundance.  We use observations of dwarf spheroidal galaxies by the Fermi Gamma-Ray Telescope and observations of the Galactic Center by the High Energy Stereoscopic System to constrain these kination scenarios.  Combining the unitarity constraint on $\langle \sigma v \rangle$ with these observational constraints sets a lower limit on the temperature at which the Universe can become radiation dominated following a period of kination if ${\langle \sigma v \rangle > 3\times 10^{-31} \sigunits}$.  This lower limit is between ${0.05 \, \mathrm{GeV}}$ and ${1 \, \mathrm{GeV}}$, depending on the dark matter annihilation channel.
\end{abstract}

\maketitle

\section{Introduction}
\label{sec:Intro}

The expansion history of the Universe before Big Bang Nucleosynthesis (BBN) is uncertain.  The fact that the primordial curvature perturbation spectrum is almost scale invariant strongly suggests that shortly after the Big Bang, the Universe experienced a period of inflation \cite{Guth:1980, Linde:1981, Albrecht:1982}.  The energy scale of inflation is not known, but it is generally assumed to be greater than ${10^{10} \, \mathrm{GeV}}$.  The successful BBN prediction of the abundances of light elements only requires that the Universe be radiation dominated at a temperature of ${3 \, \mathrm{MeV}}$ \cite{Kawasaki:1999, Kawasaki:2000, Hannestad:2004, Ichikawa:2005, Ichikawa:2006}.  Thus, there is a gap in the cosmological record between the theorized energy scale of inflation and ${3 \, \mathrm{MeV}}$.

In the simplest model, inflation is powered by a scalar field defined as the inflaton, and the Universe becomes radiation dominated when the inflaton decays into relativistic particles \cite{Kofman:1994, Kofman:1997, Allahverdi:2010}.  Another possibility is that a different scalar field dominates the Universe after inflation \cite{Kane:2015}.  If either of these scalar fields oscillates around the minimum of their potential, it behaves like a pressureless fluid and the Universe would be in an early-matter-dominated era \cite{Turner:1983, Chung:1998, Kamionkowski:1990, Giudice:2000, Fornengo:2002, Pallis:2004, Gondolo:2006, Gelmini:2006, Drewes:2014, Kumar:2015, Erickcek:2015, Kane:2015}.  An alternative scenario is that a fast-rolling scalar field (a kinaton) dominates the energy density of the Universe prior to the onset of radiation domination.  When the kinaton's energy density is dominant, the Universe is said to be in a period of kination \cite{Spokoiny:1993, Joyce:1996, Ferreira:1997}.  Kination was initially proposed as an inflationary model that does not require the complete conversion of the false vacuum energy into radiation to initiate the onset of radiation domination \cite{Spokoiny:1993}.  Kination also facilitates baryogenesis; if the electroweak phase transition occurs during kination, then baryogenesis is possible during a second-order phase transition \cite{Joyce:1996}.  Finally, if the kinaton's potential energy becomes dominant at very late times, it can accelerate the expansion of the Universe and mimic the effects of a cosmological constant \cite{Ferreira:1997, Peebles:1998, Dimopoulos:2001, Dimopoulos:2002Curvaton, Chung:2007}.

The uncertainties in the thermal history of the Universe prior to BBN limit our understanding of the origins of dark matter \cite{Kamionkowski:1990, Giudice:2000, Gelmini:2006, Drees:2006, Grin:2007, Watson:2009, Kane:2015, Erickcek:2015, Beniwal:2017}.  We study the effects of kination on the thermal production of dark matter.  We derive analytic expressions for the dark matter relic abundance generated during kination and confirm that our analytic results match the numeric solutions to the Boltzmann equation.  Our relic abundance expressions depend on the dark matter mass $m_\chi$, the velocity-averaged dark matter annihilation cross section $\langle \sigma v \rangle$, and the temperature at which the Universe becomes radiation dominated, $T_\RH$.  Our analytic expressions allow us to solve for the $\langle \sigma v \rangle$ values that will generate the observed dark matter abundance.  We determine that in order to achieve the observed dark matter abundance, kination models in which dark matter reaches thermal equilibrium require $\langle \sigma v \rangle$ values that would underproduce dark matter during radiation domination.  In contrast, kination models in which dark matter does not reach thermal equilibrium require $\langle \sigma v \rangle$ values that would overproduce dark matter during radiation domination.  Using the most recent constraints on $m_\chi$ and $\langle \sigma v \rangle $ from Fermi-LAT PASS-8 observations of dwarf spheroidal galaxies~\cite{Fermi:Constraints} and High Energy Stereoscopic System (H.E.S.S.) observations of the Galactic Center~\cite{HESS:Constraints}, we constrain $T_{\RH}$ for kination scenarios where dark matter reaches thermal equilibrium.

Prior investigations of dark matter production during kination have focused on specific kinaton potentials.  References \cite{Profumo:2003, Pallis:2005, Pallis:2nd2005, Pallis:2006, Gomez:2008} investigated how the relic abundance of dark matter is affected if the kinaton has an exponential potential, while Refs.~\cite{Lola:2009, Pallis:2009} studied kination models where the kinaton has an inverse power-law potential.  While these prior works did place constraints on dark matter parameters, those constraints were dependent on the specified kinaton potential.  Our relic abundance expressions are independent of the kinaton potential, and our constraints on $m_\chi$, $T_{\RH}$, and $\langle \sigma v \rangle$ are applicable to all kination scenarios in which dark matter is a thermal relic.  Furthermore, improvements in the observational constraints on $\langle \sigma v \rangle$ over the past six years allow us to place tighter constraints than previous works.  We determine that kination scenarios in which dark matter reaches thermal equilibrium have a minimum allowed reheat temperature between ${0.05 \, \mathrm{GeV}}$ and ${1 \, \mathrm{GeV}}$, depending on the dark matter annihilation channel.

In Section \ref{sec:Background}, we discuss the evolution equations that govern the thermal production of dark matter during kination.  In Sections \ref{sec:FreezeOut} and \ref{sec:FreezeIn}, we present analytic derivations of the dark matter relic abundance for dark matter that does and does not reach thermal equilibrium.  In Section \ref{sec:Constraints}, we use observational data from Fermi-LAT and H.E.S.S. to constrain $m_\chi$, $T_{\RH}$, and $\langle \sigma v \rangle$.  In Section \ref{sec:end}, we summarize our results.  Natural units $(\hbar = c=k_B=1)$ are used throughout this work.

\section{Thermal Dark Matter During Kination}
\label{sec:Kination}

\subsection{Kinaton Cosmology}
\label{sec:Background}

The scenario we consider consists of a fast-rolling scalar field (the kinaton) that dominates the energy density of the Universe prior to BBN.  The kinaton's energy density is dominated by its kinetic energy, meaning that the kinaton's energy density equals its pressure and that the equation of state parameter is $w=1$.  Therefore, the kinaton's energy density scales as $a^{-6}$, where $a$ is the scale factor, and will eventually become subdominant to radiation, whose energy density scales as $a^{-4}$.  Reheating is defined as the point at which the radiation energy density becomes the dominant component of the Universe.  It is important to note, however, that during kination, the temperature of the radiation bath is higher than the temperature at reheating.  Therefore, it is possible to thermally produce dark matter prior to the onset of radiation domination.

We consider three energy density components during kination: dark matter, radiation, and the kinaton.  The evolution of these energy densities are governed by three free parameters: the dark matter mass $m_{\chi}$, the reheat temperature $T_{\RH}$, and the velocity-averaged dark matter annihilation cross section $\langle \sigma v \rangle$.  Throughout this work, we assume \textit{s}-wave dark matter annihilation.  In kination scenarios, $T_\RH$ is the temperature at which the radiation energy density equals the kinaton energy density.  We assume that the kinaton does not decay nor interact with dark matter or radiation (see Ref.~\cite{Pallis:2nd2005} for an analysis of decaying kinaton cosmologies).  Radiation and dark matter on the other hand are thermally coupled via pair production and annihilation.  Therefore, the equations for the energy density of the scalar field $\rho_{\phi}$, the radiation energy density $\rho_r$, and the dark matter number density $n_\chi$ are
\begin{subequations}
\begin{align}
& \frac{d}{dt} \rho_{\phi} = -6H\rho_{\phi}, \\
& \frac{d}{dt}n_{\chi} = -3Hn_{\chi} - \langle \sigma v \rangle (n_{\chi}^2 - n_{\chi,\eq}^2), \\
& \frac{d}{dt}\rho_r = -4H\rho_r + \langle \sigma v \rangle E_{\chi} (n_{\chi}^2 - n_{\chi,\eq}^2),
\end{align}
\label{eq:Boltz}%
\end{subequations}
where ${ \langle E_{\chi} \rangle = \rho_{\chi}/n_{\chi}}$ is the average energy of a dark matter particle and $n_{\chi,\eq}$ is the number density of dark matter particles in thermal equilibrium.\footnote{Throughout this work, we assume that the dark matter is composed of Majorana particles, and therefore ${\chi = \bar{\chi}}$.}  For a dark matter particle with mass $m_{\chi}$ and internal degrees of freedom $g_{\chi}$ within a thermal bath of temperature $T$,
\begin{figure}
\centering\includegraphics[width=3.4in]{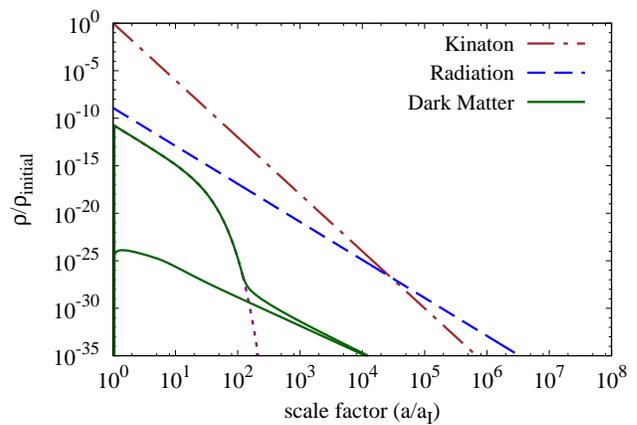}
\caption{The density evolution of the kinaton, radiation, and dark matter.  In this figure, ${m_\chi = 10^4 \, \mathrm{GeV}}$, and reheating occurs when ${a/a_I=2.7\times 10^4}$; the reheat temperature is ${2\,\mathrm{GeV}}$.  The two solid curves show the evolution of $\rho_{\chi}$ for the two values of $\langle \sigma v \rangle$ that produce the observed dark matter density: ${\Omega_{\chi} h^2 = 0.12}$ \cite{Planck:2015}.  The top solid curve corresponds to the freeze-out scenario with ${\langle \sigma v \rangle = 7.5 \times 10^{-25} \sigunits}$, whereas the bottom solid curve corresponds to the freeze-in scenario with ${\langle \sigma v \rangle = 6.7\times10^{-46} \sigunits}$.  The dotted line shows the equilibrium dark matter density, ${\rho_{\chi,\mathrm{eq}} = \langle E_{\chi} \rangle n_{\chi,\mathrm{eq}}}$.}
\label{Fig:DensityEvolution}
\end{figure}
\begin{align}
n_{\chi,\eq} = \frac{g_{\chi}}{2\pi^2} \int_{m_{\chi}}^{\infty} \frac{\sqrt{E^2 - m_{\chi}^2}}{e^{E/T} +1} E \, \mathrm{d}E .
\label{eq:Equilibrium}
\end{align}
When evaluating the average energy of a dark matter particle, we make the approximation that ${\langle E_{\chi} \rangle \simeq \sqrt{m_{\chi}^2 + (3.151 \, T)^2}}$, which matches ${\rho_{\chi}/n_{\chi}}$ to within $10\%$.

Figure \ref{Fig:DensityEvolution} shows the evolution of the kinaton, radiation, and dark matter densities obtained by numerically solving Eq.~(\ref{eq:Boltz}).  Initially, the kinaton's energy density is dominant, but since it scales away more quickly than the radiation energy density, it eventually becomes subdominant.  Figure \ref{Fig:DensityEvolution} shows that the radiation energy density scales as $a^{-4}$ and is unaffected by dark matter annihilation or pair production.  Since $\rho_r$ and $\rho_\phi$ evolve independently of $n_\chi$, we can solve for their evolution analytically.  We then use these solutions to numerically solve Eq.~(\ref{eq:Boltz}b) and calculate the dark matter relic abundance.

To accurately describe the evolution of $\rho_r$, we need to take into account the energy injection that occurs when Standard Model particles become nonrelativistic.  When a particle species becomes nonrelativistic, its entropy is transferred to the remaining relativistic particles.  Entropy is conserved during kination; therefore, the universal entropy $sa^3$ must remain constant, where $s$ is the entropy density: ${s \equiv (2\pi^2/45) \, T^3 g_{*s}(T)}$, and $g_{*s}(T)$ is the effective number of degrees of freedom that contribute to the entropy density.  Due to the conservation of entropy, radiation cools during kination according to the same proportionality as during radiation domination: ${T \propto g_{*s}(T)^{-1/3} \, a^{-1}}$.

To evaluate the temperature of the radiation bath, we set a maximum temperature of $T_{\mathrm{MAX}}$ at which ${\rho_{\chi} =0}$.  We set ${T_{\mathrm{MAX}} = 8m_{\chi}}$ to ensure that if the dark matter is capable of reaching thermal equilibrium, it will have adequate time to do so.  If the dark matter cannot reach thermal equilibrium, setting ${T_{\mathrm{MAX}} = 8m_{\chi}}$ ensures there will be enough time for maximal pair production.  Therefore, the dark matter relic abundance will not be sensitive to $T_{\mathrm{MAX}}$.  Using $T_{\mathrm{MAX}}$, we construct an expression for the temperature evolution during kination that accounts for changes in $g_{*s}(T)$:
\begin{align}
T = T_{\mathrm{MAX}} \left[\frac{g_{*s}(T_\mathrm{MAX})}{g_{*s}(T)}  \right]^{1/3} \, \frac{a_I}{a},
\label{eq:Temperature}
\end{align}
where $a_I$ is the scale factor value when ${T=T_{\mathrm{MAX}}}$.

The final step in evaluating $\rho_r$ is to connect Eq.~(\ref{eq:Temperature}) and the definition of $\rho_r$.  The radiation energy density is $\rho_r \equiv (\pi^2/30) \, g_*(T) \, T^4$, where $g_{*}(T)$ is the number of relativistic degrees of freedom at temperature $T$.  Using this definition of $\rho_r$ and Eq.~(\ref{eq:Temperature}), we see that the evolution of $\rho_r$ during kination is
\begin{align}
\rho_r = \frac{\pi^2}{30} \, g_*(T) \, T_\mathrm{MAX}^4 \left[ \frac{g_{*s}(T_\mathrm{MAX})}{g_{*s}(T)} \right]^{4/3} \left(\frac{a_I}{a}  \right)^4.
\label{eq:RadiationEnergyDensity}
\end{align}

Next, we analytically solve for $\rho_\phi$.  Solving Eq.\!~(\ref{eq:Boltz}a) yields ${\rho_\phi = \rho_{\phi,I} \, (a_I/a)^6}$, where $\rho_{\phi,I}$ is $\rho_{\phi}$ when $a=a_I$.  By defining $a_{\RH}$ as the scale factor value at the onset of radiation domination we see that $\rho_\phi$ evaluated at reheating equals ${\rho_{\phi,I} \, (a_I/a_\RH)^6}$.  Using Eq.~(\ref{eq:RadiationEnergyDensity}), we can evaluate $\rho_r$ at reheating.  Considering that at reheating ${\rho_\phi = \rho_r}$, this implies that
\begin{align}
\rho_{\phi,I} = \frac{\pi^2}{30} \, g_*(T_\RH) \, T_\mathrm{MAX}^4 \left[ \frac{g_{*s}(T_\mathrm{MAX})}{g_{*s}(T_\RH)} \right]^{4/3} \left(\frac{a_\RH}{a_I}  \right)^2.
\label{eq:InitialScalarEnergyDensity}
\end{align}
Using Eq.~(\ref{eq:Temperature}) to relate $a_\RH$ to $T_\RH$, we obtain the evolution of $\rho_\phi$ during kination:
\begin{align}
\rho_{\phi} = \frac{\pi^2}{30} \left[ \frac{T_\mathrm{MAX}^3}{T_\RH} \right]^2 \left[\frac{g_{*s}(T_\mathrm{MAX})}{g_{*s}(T_\RH)}   \right]^2 g_*(T_\RH) \left(\frac{a_I}{a}  \right)^6.
\label{eq:ScalarEnergyDensity}
\end{align}

Now that we have obtained expressions for $T(a)$, $\rho_r(a)$ and $\rho_\phi(a)$, we have the necessary components to numerically solve Eq.~(\ref{eq:Boltz}b) for $n_\chi(a)$, as shown in Figure \ref{Fig:FreezeoutDarkMatter}.  Figure \ref{Fig:OmegaDM} shows the dark matter relic abundance as a function of $\langle \sigma v \rangle$ for several values of $T_\RH$ and $m_\chi$.  For small $\langle \sigma v \rangle$ values, the dark matter cannot reach thermal equilibrium, and Figure \ref{Fig:OmegaDM} shows that as $\langle \sigma v \rangle$ increases the dark matter relic abundance increases.  Once $\langle \sigma v \rangle$ becomes large enough, pair production will bring $n_\chi$ up to its thermal equilibrium value.  If dark matter reaches thermal equilibrium, we see from Figure \ref{Fig:OmegaDM} that as $\langle \sigma v \rangle$ increases, the dark matter relic abundance decreases.  In the following sections, we derive analytic expressions for the dark matter relic abundance generated during kination and analyze how the relic abundance is influenced by $T_\RH$.

\subsection{Freeze-Out}
\label{sec:FreezeOut}

\begin{figure}
\centering\includegraphics[width=3.4in]{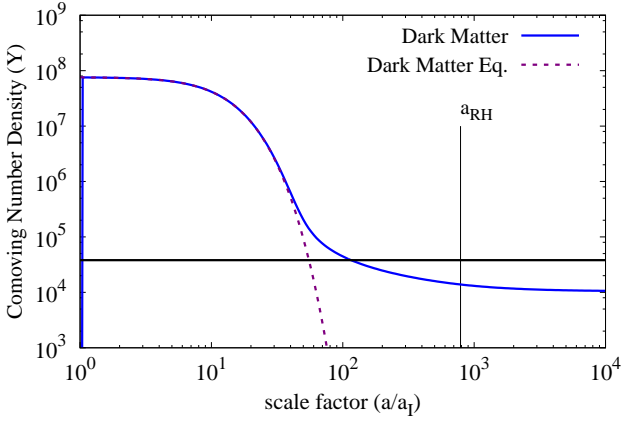}
\caption{The evolution of the comoving dark matter number density and equilibrium number density with ${T_{\RH} = 20 \, \mathrm{GeV}}$, ${m_{\chi} = 3000 \, \mathrm{GeV}}$, and ${\langle \sigma v \rangle = 10^{-32} \sigunits}$.  The vertical line represents the point of reheating at ${a_{\RH} /a_I = 800}$.  The solid horizontal line shows the comoving number density at the point of freeze-out solved from ${H{\left(T_\mathrm{F} \right)} = \langle \sigma v \rangle n_{\chi,\eq}}$.  This figure demonstrates that dark matter annihilations after freeze-out significantly decrease the dark matter number density.}
\label{Fig:FreezeoutDarkMatter}
\end{figure}

If $\langle \sigma v \rangle$ is sufficiently large, then pair production brings dark matter into thermal equilibrium: ${n_{\chi} = n_{\chi,\eq}}$, as defined in Eq.~(\ref{eq:Equilibrium}).  Once ${H \simeq \langle \sigma v \rangle n_{\chi,\eq}}$, the dark matter deviates from equilibrium and ``freezes out".  If dark matter freezes out during radiation domination, nearly all dark matter annihilations cease at freeze-out.  However, if dark matter freezes out during kination, we see from Figure \ref{Fig:FreezeoutDarkMatter} that we need to take dark matter annihilations between the time of freeze-out and reheating into account to get an accurate relic abundance.

To analytically solve for the evolution of the dark matter number density between freeze-out and reheating, we define a dimensionless comoving number density ${Y \equiv n_{\chi} (a/a_I)^3 \, T_{\RH}^{-3}}$.  Equation~(\ref{eq:Boltz}$\mathrm{b}$) is rewritten as
\begin{align}
\frac{dY}{da} = \langle \sigma v \rangle \frac{T_{\RH}^3 \, a_I^3}{Ha^4} (Y_{\eq}^2 - Y^2).
\label{eq:Dimensionless}
\end{align}
After the dark matter freezes out, ${Y^2 \gg Y_{\eq}^2}$.  Since during kination ${H = H(a_I) [a_I/a]^{3}}$, we simplify Eq.~(\ref{eq:Dimensionless}) to
\begin{align}
\frac{dY}{da} = \frac{-\lambda_{\mathrm{KD}}}{a} \, Y^2,
\label{eq:ReducedBoltzEqn}
\end{align}
where ${\lambda_{\mathrm{KD}} = T_{\RH}^3 \langle \sigma v \rangle / H(a_I)}$.  Integrating Eq.~(\ref{eq:ReducedBoltzEqn}) from freeze-out to reheating yields
\begin{align}
\frac{1}{Y_\mathrm{F}} - \frac{1}{Y_{\RH}} = -\lambda_{\mathrm{KD}} \, \ln{\frac{a_{\RH}}{a_{\mathrm{F}}}} ,
\label{eq:SolvedBoltzEqn}
\end{align}
where $Y_{\mathrm{F}}$ and $Y_{\mathrm{RH}}$ are the comoving dark matter number densities at freeze-out and reheating.  Therefore, if freeze-out occurs during kination, the dark matter comoving number density experiences a logarithmic decrease between freeze-out and reheating.

To evaluate the current dark matter density we need to reevaluate Eq.~(\ref{eq:Dimensionless}) during radiation domination and solve for $Y$ at some late time.  During radiation domination ${H = H(a_\RH) [a_\RH/a]^{2}}$, and by defining ${\lambda_{\mathrm{RD}} = [T_{\RH}^3 \langle \sigma v \rangle / H(a_\RH)]\times[a_I^3/a_\RH^2]}$, Eq.~(\ref{eq:Dimensionless}) simplifies to
\begin{align}
\frac{dY}{da} = \frac{-\lambda_{\mathrm{RD}}}{a^2} \, Y^2.
\label{eq:ReducedBoltzEqnRD}
\end{align}
Solving Eq.~(\ref{eq:ReducedBoltzEqnRD}) from reheating to a very late time yields
\begin{align}
\frac{1}{Y_\mathrm{RH}} - \frac{1}{Y_{\mathrm{LT}}} = -\lambda_{\mathrm{RD}} \left(\frac{1}{a_\mathrm{RH}}\right),
\label{eq:SimplifySolRD}
\end{align}
where $Y_{\mathrm{LT}}$ is the comoving dark matter number density at some late time $(a=a_\mathrm{LT})$.  To obtain Eq.~(\ref{eq:SimplifySolRD}), we use the fact that ${a_{\mathrm{LT}} \gg a_{\mathrm{RH}}}$.  Therefore, if dark matter freezes out during kination, $Y$ experiences a logarithmic decrease between freeze-out and reheating, after which $Y$ approaches a constant value.

Utilizing $Y_\RH$ from Eq.~(\ref{eq:SolvedBoltzEqn}) and rewriting Eq.~(\ref{eq:SimplifySolRD}) in terms of the dark matter number density yields
\begin{figure*}[t]
 \centering
\begin{minipage}{0.5\textwidth}
\centering
 \resizebox{3.4in}{!}
 {
      \includegraphics{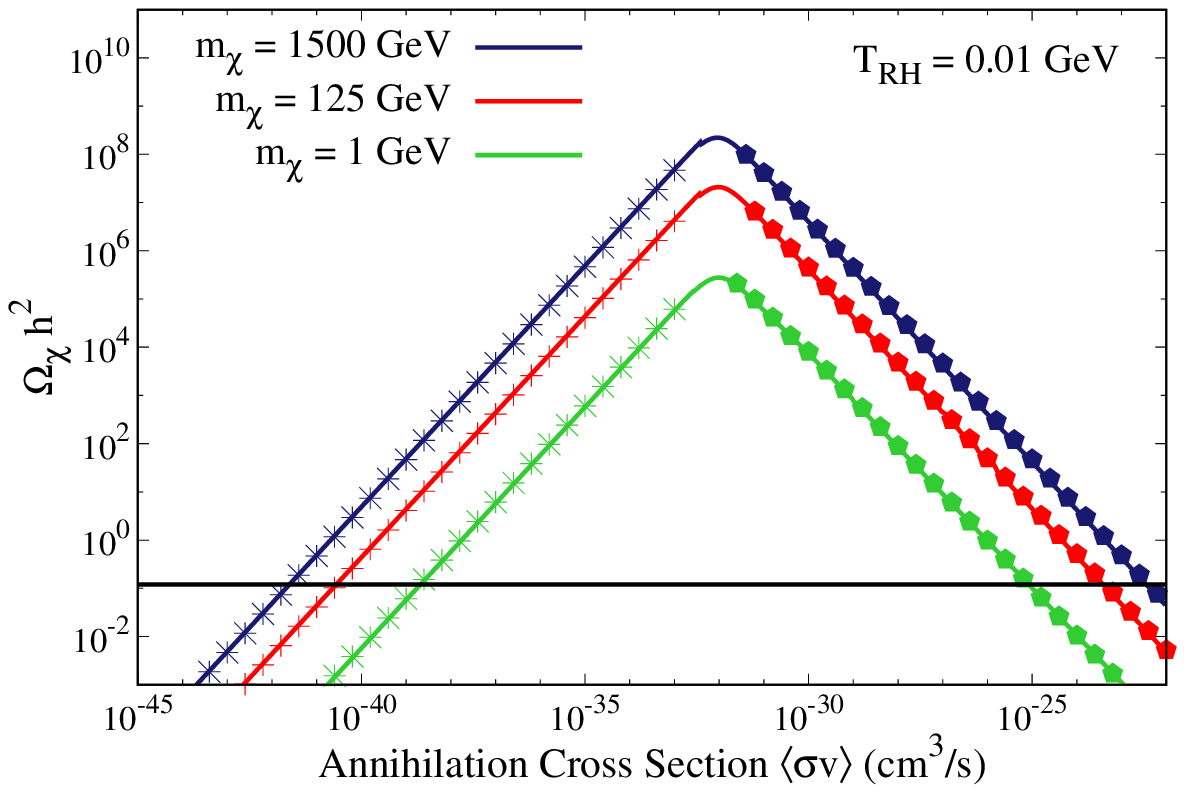}
 }
\end{minipage}%
\begin{minipage}{0.5\textwidth}
\centering
 \resizebox{3.4in}{!}
{
      \includegraphics{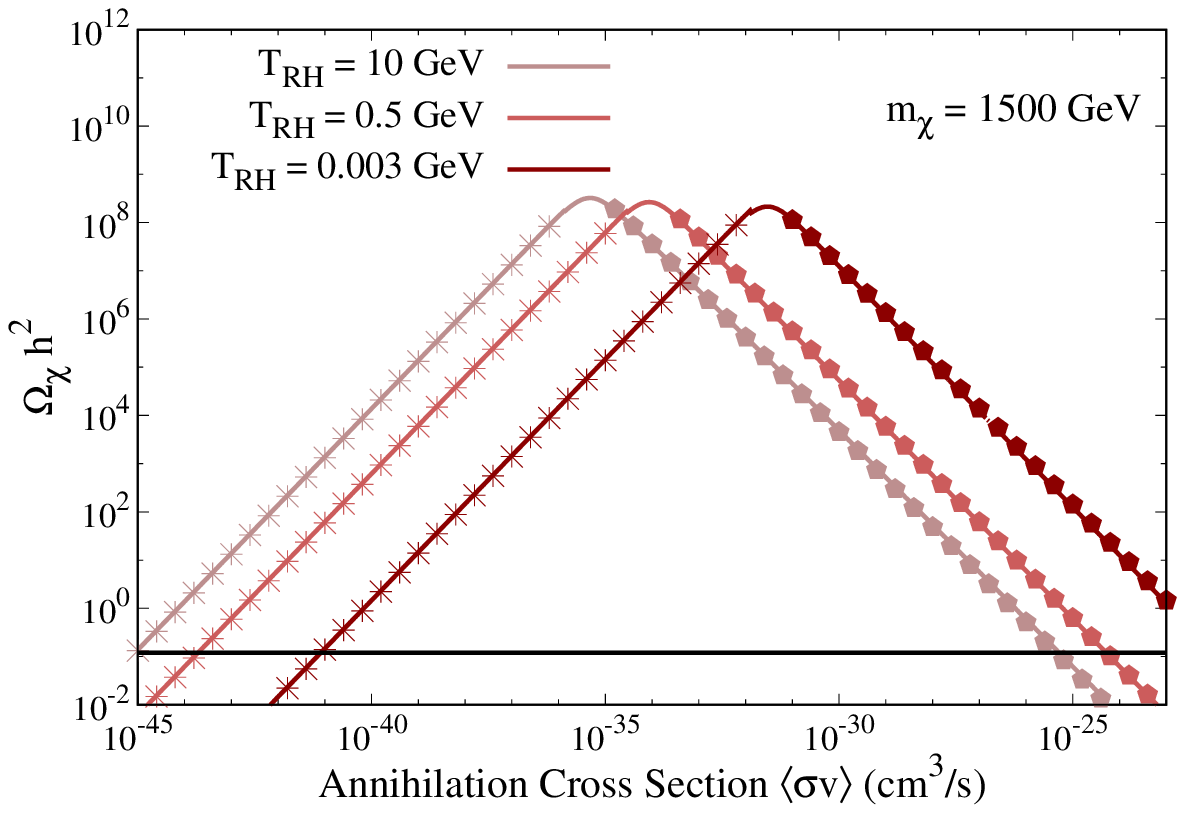}
 }
\end{minipage}%
\caption{The observed dark matter abundance ${\Omega_{\chi} h^2}$ as a function of the dark matter velocity-averaged annihilation cross section $\langle \sigma v \rangle$.  Dark matter freezes in at small $\langle \sigma v \rangle$ where ${\Omega_{\chi} h^2 \propto \langle \sigma v \rangle}$, whereas dark matter freezes out at large $\langle \sigma v \rangle$ where ${\Omega_{\chi} h^2 \propto \langle \sigma v \rangle^{-1}}$.  In the left panel we see that decreasing $m_{\chi}$ decreases $\Omega_{\chi} h^2$ for both cases.  In the right panel we see that decreasing $T_{\RH}$ decreases ${\Omega_{\chi} h^2}$ for the freeze-in case but increases ${\Omega_{\chi} h^2}$ for the freeze-out case.  In both panels the solid curves represent the numerical solutions to Eq.~(\ref{eq:Boltz}), while the symbols represent the analytic approximations represented by Eqs.~(\ref{eq:FreezeoutRelicAbundance}) and~(\ref{eq:FreezeInRelicAbundance}).  The solid black line represents the Planck measurement for the observed dark matter abundance, ${\Omega_{\chi} h^2 = 0.12}$ \cite{Planck:2015}.}
\label{Fig:OmegaDM}
\end{figure*}
\begin{align}
n_{\chi,\mathrm{LT}} =& \,\, \left[\frac{\langle \sigma v \rangle \, a_\mathrm{LT}^3}{H(a_I) \, a_I^3} \, \left(\ln \left[\frac{a_{\RH}}{a_\mathrm{F}} \right]+1\right)+\frac{a_\mathrm{LT}^3}{n_{\chi,\mathrm{F}} \, \, a_\mathrm{F}^3} \right]^{-1}.
\label{eq:ReheatingNumberDensityFirst}
\end{align}
We wish to express Eq.~(\ref{eq:ReheatingNumberDensityFirst}) in terms of our free parameters $m_{\chi}$, $T_{\RH}$, and $\langle \sigma v \rangle$.  We can express $H(a_I)$ and $a_{\RH}$ in terms of $T_{\RH}$ and $T_{\mathrm{MAX}}$ using Eqs.~(\ref{eq:Temperature}) and~(\ref{eq:InitialScalarEnergyDensity}).  In addition, since ${n_{\chi, \mathrm{F}} \simeq n_{\chi,\mathrm{eq}}}$, ${n_{\chi,\mathrm{F}} \simeq H{\left(T_\mathrm{F} \right)}/ \langle \sigma v \rangle}$.  If freeze-out occurs during kination, ${H^2 \simeq \left(8 \pi G/3 \right) \rho_\phi}$; combining this with  Eqs.~(\ref{eq:Temperature}) and~(\ref{eq:ScalarEnergyDensity}) allows us to solve for $H{\left(T_\mathrm{F} \right)}$:
\begin{align}
{H\!\left(T_\mathrm{F}\right)} = \,\, \left( \frac{4\pi^3}{45} \right)^{1/2} \left( \frac{T_\mathrm{F}^3}{m_\mathrm{pl} \, T_\RH}  \right) \left( \frac{g_{*s\mathrm{F}}}{g_{*s\mathrm{RH}}}  \right) g_{*\mathrm{RH}}^{1/2},
\label{eq:HubbleTemperature}
\end{align}
where ${g_{*s\RH} = g_{*s}(T_{\RH})}$ and ${g_{*\RH} = g_{*}(T_{\RH})}$.  These relations allow us to rewrite Eq.~(\ref{eq:ReheatingNumberDensityFirst}) as
\begin{align}
n_{\chi,\mathrm{LT}} =& \,\, \left(\frac{4\pi^3}{45} \right)^{1/2} \frac{T_{\mathrm{LT}}^3 \, g_{*s\mathrm{LT}} \, g_{*\mathrm{RH}}^{1/2}}{\langle \sigma v \rangle \, \mpl \, T_\RH \, g_{*s\RH}} \\
&\times \left(\ln\left[\frac{T_\mathrm{F}}{T_{\RH}} \left(\frac{g_{*s\mathrm{F}}}{g_{*s\RH}}\right)^{1/3}\right] + 2\right)^{-1}, \nonumber
\label{eq:ReheatingNumberDensity}
\end{align}
where $T_\mathrm{F}$ is obtained by numerically solving ${H(T_\mathrm{F}) = \langle \sigma v \rangle n_{\chi,\mathrm{eq}}}$.  For kination scenarios, ${m_\chi / T_{\mathrm{F}}}$ is roughly between 20 and 30.

After $a_\mathrm{LT}$, ${n_\chi \propto a^{-3}}$, which allows us to relate the dark matter density at $a_\mathrm{LT}$ to today:
\begin{align}
\rho_{\chi,0} = \rho_{\chi,\mathrm{LT}} \left(\frac{a_{\mathrm{LT}}}{a_0}\right)^3 = \rho_{\chi,\mathrm{LT}} \left(\frac{T_0}{T_{\mathrm{LT}}}\right)^3 \left(\frac{g_{*s0}}{g_{*s\mathrm{LT}}}\right),
\end{align}
where $T_0$ is the radiation temperature today and ${g_{*s0} = 3.91}$.  Bringing all of the previous components together and scaling our analytic expression by a factor of $1.22$, thereby ensuring that it matches the numeric solution of Eq.~(\ref{eq:Boltz}b) within 20\% for ${m_\chi / T_\RH > 100}$, we obtain an analytic expression for the freeze-out dark matter relic abundance:
\begin{align}
\Omega_{\chi} h^2 =& \,\, 6.06 \left(\frac{3\times10^{-26} \sigunits}{\langle \sigma v \rangle}\right) \frac{g_{*\mathrm{RH}}^{1/2}}{g_{*s\mathrm{RH}}}
\left(\frac{m_{\chi}/T_{\RH}}{150}\right) \nonumber \\
&\times \left(\ln\left[\frac{T_\F/T_\RH}{10} \left(\frac{g_{*s\F}}{g_{*s\RH}}\right)^{1/3}\right] + 4.3\right)^{-1}.
\label{eq:FreezeoutRelicAbundance}
\end{align}
Equation~(\ref{eq:FreezeoutRelicAbundance}) indicates that decreasing $T_{\RH}$ increases the relic abundance.  Decreasing $T_{\RH}$ requires increasing the kinaton energy density, which increases the Hubble parameter during kination.  Since ${n_{\chi,\mathrm{F}} \simeq H(T_{\mathrm{F}})/\langle \sigma v \rangle}$, increasing the Hubble parameter increases the dark matter number density at freeze-out and thus increases the relic abundance.  Furthermore, in our calculation of $Y_{\RH}$ we showed that dark matter annihilations do not cease during kination.  As a result, Eq.~(\ref{eq:FreezeoutRelicAbundance}) includes an inverse logarithmic term that depends on the ratio ${T_{\mathrm{F}}/T_\RH}$.

Figure \ref{Fig:OmegaDM} shows the dark matter relic abundance as a function of $\langle \sigma v \rangle$ for several values of $T_\RH$ and $m_\chi$.  In Figure \ref{Fig:OmegaDM} we see that for sufficiently large $\langle \sigma v \rangle$ the freeze-out dark matter relic abundances from Eq.~(\ref{eq:FreezeoutRelicAbundance}), represented by the circular symbols, match the numeric solutions to Eq.~(\ref{eq:Boltz}$\mathrm{b}$), represented by the curves.  We can solve for the minimum $\langle \sigma v \rangle$ that will result in the dark matter reaching thermal equilibrium.  If dark matter freezes out, ${H{\left(T_\mathrm{F} \right)} = \langle \sigma v \rangle n_{\chi,\eq}}$, and we can rewrite this equation in terms of a new variable $x$, where $x \equiv m_{\chi}/T_\mathrm{F}$.  Assuming that dark matter is nonrelativistic, ${H{\left(T_\mathrm{F} \right)} = \langle \sigma v \rangle n_{\chi,\eq}}$ can be rewritten as ${x^{-3/2} e^{x} g_{*s}\left(m_\chi /x\right) = constant \times \langle \sigma v \rangle}$.  The left-hand side of this equation has a minimum value near ${x \sim 1.5}$ which implies that there is a minimum $\langle \sigma v \rangle$ for which a solution will exist.  This lower bound on $\langle \sigma v \rangle$ is
\begin{align}
\langle \sigma v \rangle > & \,\, 9.37 \times 10^{-33} \sigunits \left(\frac{2}{g_\chi} \right) \nonumber \\
&\times  \left( \frac{3 \, \mathrm{MeV}}{T_{\RH}} \right) \left( \frac{g_{*\RH}^{1/2}}{g_{*s\RH}}\right) g_{*s}(m_\chi/1.5)  .
\label{eq:CrossSectionBeginningFreezeOut}
\end{align}

The horizontal line in Figure \ref{Fig:OmegaDM} represents the Planck measurement of the observed dark matter abundance.  To reproduce the observed dark matter abundance, freeze-out cases during kination require larger $\langle \sigma v \rangle$ than that required if freeze-out occurs during radiation domination.  This comes from the fact that at a given temperature the Hubble parameter during kination is always higher than it is during radiation domination, which causes freeze-out to occur earlier.  In order to compensate for the earlier freeze-out and reproduce the observed dark matter abundance, freeze-out scenarios during kination require ${\langle \sigma v \rangle > 3\times 10^{-26} \sigunits}$.  Since this lower bound on $\langle \sigma v \rangle$ is more stringent than Eq.~(\ref{eq:CrossSectionBeginningFreezeOut}), Eq.~(\ref{eq:FreezeoutRelicAbundance}) is applicable to all freeze-out scenarios that generate the observed dark matter abundance.

\subsection{Freeze-In}
\label{sec:FreezeIn}

For cross sections that violate Eq.~(\ref{eq:CrossSectionBeginningFreezeOut}), dark matter pair production is not sufficient to bring the dark matter into thermal equilibrium.  Once pair production ceases, the dark matter ``freezes in" and the comoving dark matter number density remains constant.  The first step in determining the freeze-in dark matter relic abundance is to calculate the comoving dark matter number density when pair production ceases.

In freeze-in scenarios, the dark matter number density does not reach thermal equilibrium, so ${n_{\chi} \ll n_{\chi,\eq}}$.  The dimensionless comoving dark matter equilibrium number density is ${Y_{\eq} \equiv n_{\chi,\eq} T_{\RH}^{-3} (a/a_I)^3}$, and for freeze-in scenarios ${Y_{\eq} \gg Y}$.  Therefore, Eq.~(\ref{eq:Dimensionless}) reduces to
\begin{align}
\frac{dY}{da} = \frac{\langle \sigma v \rangle \, T_{\RH}^3}{H\!\left(a_I\right) \, a } \, Y_{\eq}^2
\label{eq:FreezeInDimensionless}
\end{align}
for freeze-in scenarios during kination.

Equation~(\ref{eq:FreezeInDimensionless}) implies that $dY/da$ diverges as $a \rightarrow 0$ if $\langle \sigma v \rangle$ is independent of temperature.  The same divergence occurs if the Universe is radiation dominated during dark matter production, and it would make the freeze-in abundance of dark matter dependent on $T_\mathrm{MAX}$.  Previous analyses of the freeze-in process avoided this sensitivity to high-energy physics by assuming that  ${\langle \sigma v \rangle \propto 1/ \, T^2}$ for relativistic particles \cite{Hall:2009, Yaguna:2011}.  We take the same approach and set ${\langle \sigma v \rangle = \langle \sigma v \rangle_s (m_\chi/T)^2}$ for ${T > m_\chi}$ and ${\langle \sigma v \rangle = \langle \sigma v \rangle_s}$ for ${T < m_\chi}$, where $\langle \sigma v \rangle_s$ is the \textit{s}-wave dark matter annihilation cross section for massive particles.  With this scaling, ${dY/da \rightarrow 0}$ as ${a \rightarrow 0}$, and the production of dark matter is finite during kination even if $T_\mathrm{MAX} \rightarrow \infty.$  Figure \ref{FreezeInDimensionless} shows that $dY/da$ increases until ${a=a_*}$, which we define as the scale factor value at which pair production peaks.  The temperature at which pair production peaks is $T_* = m_{\chi}$.  Therefore, $dY/da$ reaches its maximum when ${a_*/a_{\mathrm{I}} = 8[g_{*s}(\mathrm{T_{MAX}})/g_{*s}(\mathrm{T_*})]^{1/3}}$.

Figure \ref{FreezeInDimensionless} also indicates that the integral of $dY/da$ converges; $Y$ will approach a constant value as pair production becomes less and less efficient.  However, Eq.~(\ref{eq:FreezeInDimensionless}) is only valid during kination, so it will only provide an accurate dark matter density if nearly all the pair production occurs prior to reheating.  Truncating the integration of $dY/da$ at $a_{\mathrm{PP}}$, where $T(a_\mathrm{PP}) = m_\chi/3.9$, reduces the value of $Y$ by less than 1\% compared to integrating $dY/da$ out to $a=\infty$.   Therefore, pair production has effectively halted when $T < m_{\chi}/3.9$, and we can use Eq.~(\ref{eq:FreezeInDimensionless}) to compute the relic abundance of dark matter provided that $T_\mathrm{RH} < m_{\chi}/3.9$.  Integrating Eq.~(\ref{eq:FreezeInDimensionless}) from $0$ to $a_{\mathrm{PP}}$, while taking into account the fact that $\langle \sigma v \rangle$ changes from ${\langle \sigma v \rangle_s (m_\chi/T)^2}$ to ${\langle \sigma v \rangle_s}$ at ${T = m_\chi}$, gives
\begin{align}
Y_{\mathrm{PP}} =& \,\, 3.1\times 10^{-4} \left( \frac{T_{\mathrm{MAX}} /T_\RH}{150} \right)^3 \left(\frac{T_\RH}{5 \, \mathrm{GeV}}\right) \left(\frac{g_\chi}{2} \right)^2 \nonumber \\
&\times \left( \frac{ \langle \sigma v \rangle_s}{10^{-45} \sigunits}\right) \left( \frac{g_{*s\RH}}{g_{*\RH}^{1/2}} \right) \left(\frac{g_{*s\mathrm{MAX}}}{g_{*s}^2(m_\chi)}\right).
\label{eq:EndPairProduction}
\end{align}
After pair production ends, $Y$ remains nearly constant, and thus $Y_{\mathrm{PP}} = Y_\RH$. Therefore, the dark matter density at reheating can be written as
\begin{align}
\rho_{\chi,{\RH}} = m_{\chi} \, Y_{\mathrm{PP}} \, T_{\RH}^3 \left( a_{\mathrm{I}}/a_{\RH} \right)^3.
\label{eq:FreezeInEnergyDensity}
\end{align}
Equation~(\ref{eq:Temperature}) indicates that ${\left( a_{\mathrm{I}}/a_{\RH} \right)^3} \propto T_{\mathrm{MAX}}^{-3} \, g_{*s{\mathrm{MAX}}}^{-1}$.  As a result, inserting Eq.~(\ref{eq:EndPairProduction}) into Eq.~(\ref{eq:FreezeInEnergyDensity}) seems to indicate that ${\rho_{\chi,{\RH}}}$ is independent of ${T_\mathrm{MAX}}$, but this is not generically true.  When integrating Eq.~(\ref{eq:FreezeInDimensionless}) from $a=0$ to $a_{\mathrm{PP}}$ to obtain Eq.~(\ref{eq:EndPairProduction}), we effectively integrated from ${T = \infty}$ to ${T = m_\chi/3.9}$.  However, integrating instead from ${T = 8 \, m_\chi}$ to ${T = m_\chi/3.9}$ does not significantly change the result.  Therefore, if ${T_\mathrm{MAX} \geq 8 \, m_\chi}$, the freeze-in dark matter abundance does not depend on $T_\mathrm{MAX}$.  Conversely, if ${T_\mathrm{MAX}< 8 \, m_\chi}$, ${\rho_{\chi,{\RH}}}$ will decrease as $T_\mathrm{MAX}$ decreases because maximal pair production is not reached.

After reheating, $n_\chi \propto a^{-3}$, and we can use Eq.~(15) to evolve $\rho_\chi$ from reheating to today.  Combining the previous expressions and scaling our analytic expression by a factor of $0.95$ to match the numeric solution of Eq.~(\ref{eq:Boltz}b) provides an analytic expression for the freeze-in dark matter relic abundance:
\begin{figure}
\centering\includegraphics[width=3.4in]{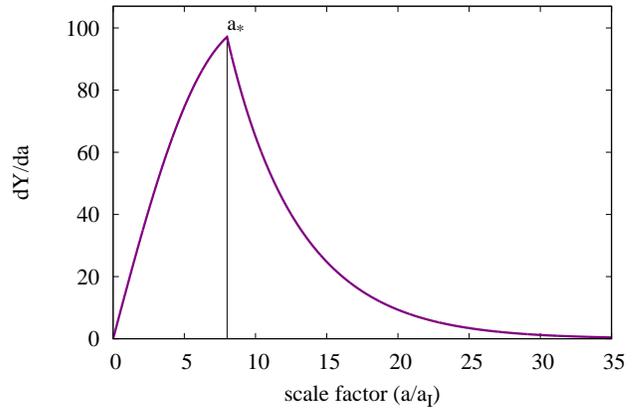}
\caption{The evolution of $dY/da$ given ${T_{\RH}=1 \, \mathrm{GeV}}$, ${m_\chi = 5\times 10^4 \, \mathrm{GeV}}$, and ${\langle \sigma v \rangle_s = 10^{-47} \sigunits}$.  The vertical line represents the scale factor at which pair production peaks, defined as $a_*$.  For kination scenarios where dark matter freezes in and $T_\mathrm{MAX} = 8\, m_\chi$, ${a_*/a_{\mathrm{I}} \simeq 8}$.}
\label{FreezeInDimensionless}
\end{figure}
\begin{align}
\Omega_{\chi} h^2 =& \, \, 1.08 \left( \frac{m_{\chi}}{1 \, \mathrm{GeV}} \right) \left(\frac{T_\RH}{100 \, \mathrm{GeV}}\right) \left(\frac{g_\chi}{2}  \right)^2
\nonumber \\
&\times \left( \frac{ \langle \sigma v \rangle_s}{10^{-45} \sigunits}\right) \left(\frac{g_{*s\RH}}{g_{*\RH}^{1/2}}   \right) g_{*s}^{-2}(m_\chi) .
\label{eq:FreezeInRelicAbundance}
\end{align}
Equation (\ref{eq:FreezeInRelicAbundance}) indicates that increasing $T_{\RH}$ leads to a larger relic abundance.  To understand how the freeze-in dark matter relic abundance relates to $T_\RH$ we need to investigate how $n_\chi$ relates to the Hubble parameter.  The connection between the Hubble parameter and $n_\chi$ stems from the cooling rate $dT/dt$.  During kination ${T\propto a^{-1}}$, and therefore ${dT/dt = -HT}$.  Rewriting ${dn/dt}$ as a function of temperature yields ${dn/dt = \left(dT/dt\right) \left(dn/dT\right)} = f(T)$, where $f(T)$ is the right-hand side of Eq.~(\ref{eq:Boltz}b).  This allows us to express $dn/dT$ as
\begin{align}
\frac{dn}{dT} = \frac{-f\left(T\right)}{HT},
\label{eq:CoolingRate}
\end{align}
which implies that ${n \propto 1/H}$.  Since increasing $T_\RH$ decreases $H$ during kination, it also decreases the cooling rate, leaving more time for pair production and thereby increasing the dark matter number density.

In order to reach the observed dark matter abundance, scenarios in which dark matter freezes in during kination require larger $\langle \sigma v \rangle_s$ values than if freeze-in occurs during radiation domination.  For example, given a dark matter mass of $100 \, \mathrm{GeV}$, a freeze-in scenario during radiation domination requires ${ \langle \sigma v \rangle_s = 10^{-47} \sigunits}$ in order for ${\Omega_{\chi} h^2 = 0.12}$ \cite{Dev:2013, Planck:2015}.  For the same $m_\chi$, a freeze-in scenario during kination with ${T_\RH = 0.033 \, \mathrm{GeV}}$  requires ${ \langle \sigma v \rangle_s = 10^{-41} \sigunits}$.  Freeze-in scenarios during kination require larger $\langle \sigma v \rangle_s$ values to generate the observed dark matter abundance because the increased cooling rate during kination leaves less time for pair production.

In Figure \ref{Fig:OmegaDM}, we see that for sufficiently small cross sections the dark matter relic abundances from Eq.~(\ref{eq:FreezeInRelicAbundance}), represented by the star symbols, match the numeric solutions to Eq.~(\ref{eq:Boltz}b), represented by the curves.  We have already discussed how freeze-in requires the dark matter particles to never reach thermal equilibrium.  For this to hold true, the dark matter number density at the end of pair production must be less than the dark matter equilibrium number density at the peak of pair production: ${Y(T_{\mathrm{PP}})<Y_{\eq}(T_*)}$.  To solve for the largest $\langle \sigma v \rangle_s$ that will not result in dark matter reaching thermal equilibrium, we approximate the equilibrium number density as being nonrelativistic:
\begin{align}
Y_{\eq}(a) = g_\chi \left(\frac{a}{a_I} \right)^3 T_{\RH}^{-3} \left(\frac{m_{\chi} T}{2\pi}\right)^{3/2} e^{-m_{\chi}/T}.
\label{eq:NonRelativisticEquilibrium}
\end{align}
Evaluating Eq.~(\ref{eq:NonRelativisticEquilibrium}) at $a_*$ and equating that to Eq.~(\ref{eq:EndPairProduction}) gives us the cross sections that will result in dark matter freezing in during kination:
\begin{align}
\langle \sigma v \rangle_s \leq & \,\, 8.5\times 10^{-34} \sigunits \left( \frac{g_{*\RH}^{1/2}}{g_{*s\RH}} \right) \nonumber \\
&\times  g_{*s}(m_\chi)  \left( \frac{2}{g_{\chi}} \right) \left(\frac{3 \, \mathrm{MeV}}{T_\RH}\right).
\label{eq:CrossSectionEndFreezeIn}
\end{align}

Figure \ref{Fig:OmegaDM} demonstrates that there is a range of $\langle \sigma v \rangle_s$ values for each reheat temperature where neither a freeze-out nor freeze-in scenario will result in the observed dark matter abundance.   The left panel of Fig.~\ref{Fig:OmegaDM} shows that, at a fixed reheat temperature, decreasing $m_\chi$ increases the freeze-in cross section and decreases the freeze-out cross section that generates the observed dark matter abundance.  However, once ${m_\chi \lesssim 3\,T_\mathrm{RH}}$, dark matter will no longer freeze in during kination, and as discussed in Section \ref{sec:FreezeOut}, $\langle \sigma v \rangle > 3\times 10^{-26} \sigunits$ is required to generate the observed dark matter abundance if dark matter freezes out during kination.

The right panel of Fig.~\ref{Fig:OmegaDM} shows that decreasing $T_\mathrm{RH}$ increases the cross section that generates the observed dark matter abundance via the freeze-in mechanism.  Therefore, setting ${T_\RH = 3 \, \mathrm{MeV}}$ gives an upper bound on the cross sections that can generate the observed dark matter abundance in freeze-in scenarios:
\begin{align}
\langle \sigma v \rangle_s < 2.66 \times 10^{-38} \sigunits \left( \frac{0.175 \, \mathrm{GeV}}{m_\chi} \right).
\label{eq:MinForbiddenCrossSection}
\end{align}
This maximal cross section is calculated using ${g_{*s}(m_\chi) = g_{*s}(0.175\, \mathrm{GeV})}$.  When deriving Eq.~(\ref{eq:FreezeInRelicAbundance}) we assumed that $g_{*s}(T)$ was approximately constant during pair production, which implies that ${\Omega_\chi h^2 \propto g^{-2}_{*s}(m_\chi)}$.  If dark matter freezes in during kination and $m_\chi$ is less than $0.17\, \mathrm{GeV}$, then the QCD phase transition occurs before the peak of pair production.  At the QCD phase transition $g_{*s}(T)$ sharply decreases, resulting in an increase in the relic abundance as given by Eq.~(\ref{eq:FreezeInRelicAbundance}).  To compensate for the increased relic abundance, freeze-in scenarios with ${m_\chi \leq 0.17\, \mathrm{GeV}}$ require cross sections smaller than the one calculated in Eq.~(\ref{eq:MinForbiddenCrossSection}).  Therefore, the largest cross section that can generate the observed dark matter abundance in freeze-in scenarios is ${2.7\times 10^{-38} \sigunits}$.

The relic abundances from Eq.~(\ref{eq:FreezeInRelicAbundance}) are within 20\% of the solutions to Eq.~(\ref{eq:Boltz}b) for ${m_\chi > 0.17\, \mathrm{GeV}}$ and $T_\RH < m_\chi /3.9$.  If ${m_\chi \leq 0.17\, \mathrm{GeV}}$, we need to take into consideration the evolution of $g_{*s}(T)$ to accurately calculate the relic abundance.  Allowing for the evolution of $g_{*s}(T)$, $Y_{\mathrm{PP}}$ is rewritten as
\begin{align}
Y_{\mathrm{PP}} =& \,\, \left(\frac{45}{4\pi^3} \right)^{1/2} \frac{\langle \sigma v \rangle_s m_{pl}}{T_\RH^2 g^{1/2}_{*\RH}} \frac{T_\mathrm{MAX}^3}{m_\chi^6} \, g_{*s\mathrm{MAX}} \, g_{*s\RH}
\nonumber \\
&\times \left(\int_{0}^{1} n^2_{\chi,\mathrm{eq}} \, x^7 \, g^{-2}_{*s}(m_\chi /x) \, dx \, \, + \right.
\nonumber \\
&\left. \int_{1}^{x_\mathrm{PP}} n^2_{\chi,\mathrm{eq}} \, x^{5} \, g^{-2}_{*s}(m_\chi /x) \, dx \right).
\label{eq:IntegralEndPairProduction}
\end{align}
Using this expression for $Y_{\mathrm{PP}}$ and a scaling factor of $0.4$, we construct a modified relic abundance expression that takes into account the evolution of $g_{*s}(T)$.  For freeze-in scenarios with ${m_\chi \leq 0.17\, \mathrm{GeV}}$ and $T_\RH < m_\chi /3.9$, this updated expression for $Y_{\mathrm{PP}}$ brings the analytic relic abundance solutions to within $25\%$ of the numeric solution to Eq.~(\ref{eq:Boltz}b).

\section{Constraints On Kination}
\label{sec:Constraints}

To constrain kination cosmologies, we first solve Eqs.~(\ref{eq:FreezeoutRelicAbundance}) and~(\ref{eq:FreezeInRelicAbundance}) for all combinations of the variables $m_{\chi}$, $T_{\RH}$, and $\langle \sigma v \rangle$ that produce the observed dark matter abundance of ${\Omega_{\chi} h^2 = 0.12}$ \cite{Planck:2015}.  We set the minimum allowed reheat temperature to ${3 \, \mathrm{MeV}}$ to ensure that the period of kination does not alter the cosmic microwave background or the abundances of light elements \cite{Kawasaki:1999, Kawasaki:2000, Hannestad:2004, Ichikawa:2005, Ichikawa:2006}.\footnote{These constraints on the reheat temperature were derived assuming that the radiation-dominated era was preceded by an early-matter-dominated era, but we expect that similar constraints would apply to kination.}  Next, we compare our allowed parameters to current constraints on $m_\chi$ and $\langle \sigma v \rangle$ from Fermi-LAT and H.E.S.S.  Specifically, we use the Fermi-LAT PASS-8 constraints from observations of dwarf spheroidals \cite{Fermi:Constraints} and H.E.S.S. constraints from observations of the Galactic Center \cite{HESS:Constraints}.  The Fermi-LAT data covers dark matter masses ranging from ${2 \, \mathrm{GeV} \leq m_{\chi} \leq 10^4 \, \mathrm{GeV}}$, while the H.E.S.S. data covers dark matter masses ranging from ${125 \, \mathrm{GeV} \leq m_{\chi} \leq 7 \times 10^4 \, \mathrm{GeV}}$.

\begin{figure}
\centering\includegraphics[width=3.4in, height=2.8in]{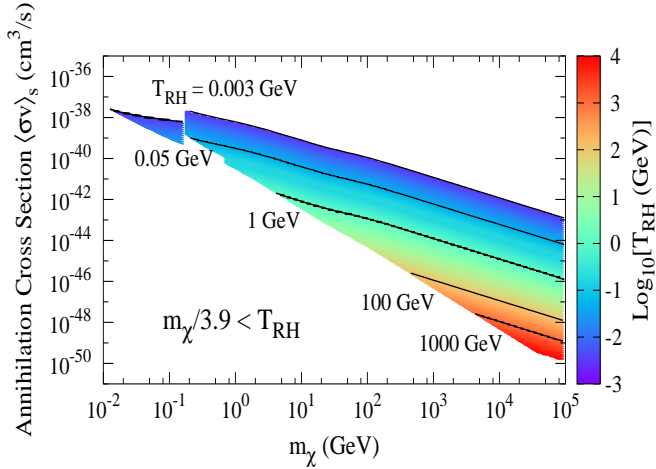}
\caption{Allowed freeze-in parameter space for $m_\chi$ and $\langle \sigma v \rangle_s$.  Equation~(\ref{eq:FreezeInRelicAbundance}) is not applicable to scenarios with ${m_{\chi}/3.9 < T_{\RH}}$ because more than 1\% of pair production occurs during radiation domination.  Dark matter produced via freeze-in requires very small annihilation cross sections in order to reach the observed dark matter abundance.  These small annihilation cross sections are not constrainable with current astrophysical observations.}
\label{NonEqConstraints}
\end{figure}

Figure \ref{NonEqConstraints} shows the allowed parameter space for $m_\chi$ and $\langle \sigma v \rangle_s$ for scenarios in which dark matter freezes in during kination.  To ensure that freeze-in occurs before the onset of radiation domination, we have restricted ourselves to ${m_{\chi}/3.9 > T_{\RH}}$.  This restriction comes from the fact that the temperature at which pair production effectively ceases is $T_\mathrm{PP} = m_\chi /3.9$.  Since the minimum reheat temperature is $3 \, \mathrm{MeV}$, we require that $m_\chi > 0.012 \, \mathrm{GeV}$ to ensure that freeze-in occurs before radiation domination.  From Eq.~(\ref{eq:MinForbiddenCrossSection}) and Figure \ref{NonEqConstraints}, we see that scenarios in which dark matter freezes in during kination require ${\langle \sigma v \rangle < 2.7\times 10^{-38} \sigunits}$.

\begin{figure}
\centering\includegraphics[width=3.4in, height=2.8in]{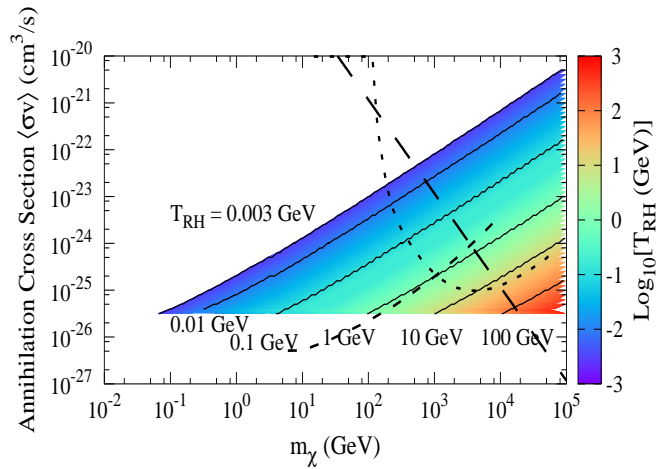}
\caption{Allowed freeze-out parameter space for $m_\chi$ and $\langle \sigma v \rangle$.  To obtain the observed dark matter abundance, scenarios with annihilation cross sections smaller than $3\times10^{-26} \sigunits$ require freeze-out to occur during radiation domination.  The short dashed line represents the H.E.S.S. \cite{HESS:Constraints} constraints for annihilation in the $b\overline{b}$ channel, and the medium dashed line is the Fermi-LAT \cite{Fermi:Constraints} constraints, also for the $b\overline{b}$ annihilation channel.  The long dashed line is the unitarity bound:\, ${\langle \sigma v \rangle \lesssim 1/m_\chi^2}$.  All of the kination scenarios above the Fermi-LAT and H.E.S.S. constraint lines are ruled out as dark matter annihilations would have already been detected by the corresponding observations.}
\label{EqConstraints}
\end{figure}

Figure \ref{EqConstraints} shows the allowed parameter space for $m_\chi$ and $\langle \sigma v \rangle$ for scenarios in which dark matter freezes out during kination.  To obtain the observed dark matter abundance, freeze-out scenarios during kination must have an annihilation cross section greater than ${3\times10^{-26} \sigunits}$.  As discussed in Section~\ref{sec:FreezeOut}, freeze-out occurs earlier during kination than during radiation domination, and to compensate, freeze-out scenarios during kination require larger annihilation cross sections to generate the same dark matter density.

In Figure \ref{EqConstraints} we include the Fermi-LAT \cite{Fermi:Constraints} and H.E.S.S. \cite{HESS:Constraints} constraints for dark matter that annihilates via the $b\overline{b}$ channel.  The Fermi-LAT bounds cover a range of dark matter masses from the mass of the bottom quark to a mass of ${10^4 \, \mathrm{GeV}}$.  The H.E.S.S. bounds add additional constraints to dark matter masses ranging from ${200 \, \mathrm{GeV}}$ to ${7 \times 10^4 \, \mathrm{GeV}}$.  Dark matter annihilation cross sections above the observational bounds are ruled out as these signals would have already been observed.  Figure \ref{EqConstraints} also includes the partial-wave unitarity bound, which requires ${\langle \sigma v \rangle \lesssim 1/m_\chi^2}$  \cite{Griest:1989, Hui:2001, Guo:2009}.  We see from Figure \ref{EqConstraints} that the unitarity bound rules out all kination scenarios with $\langle \sigma v \rangle$ values larger than $4.5\times10^{-23} \sigunits$.  In addition, if dark matter annihilates via the $b\overline{b}$ channel, Fermi-LAT and H.E.S.S. observations constrain $\langle \sigma v \rangle$ to be less than $2 \times 10^{-25} \sigunits$ and $T_\RH$ to be greater than $1 \, \mathrm{GeV}$.

\begin{figure*}[t]
 \centering
\begin{subfigure}
\centering
 \includegraphics[width=.45\textwidth]{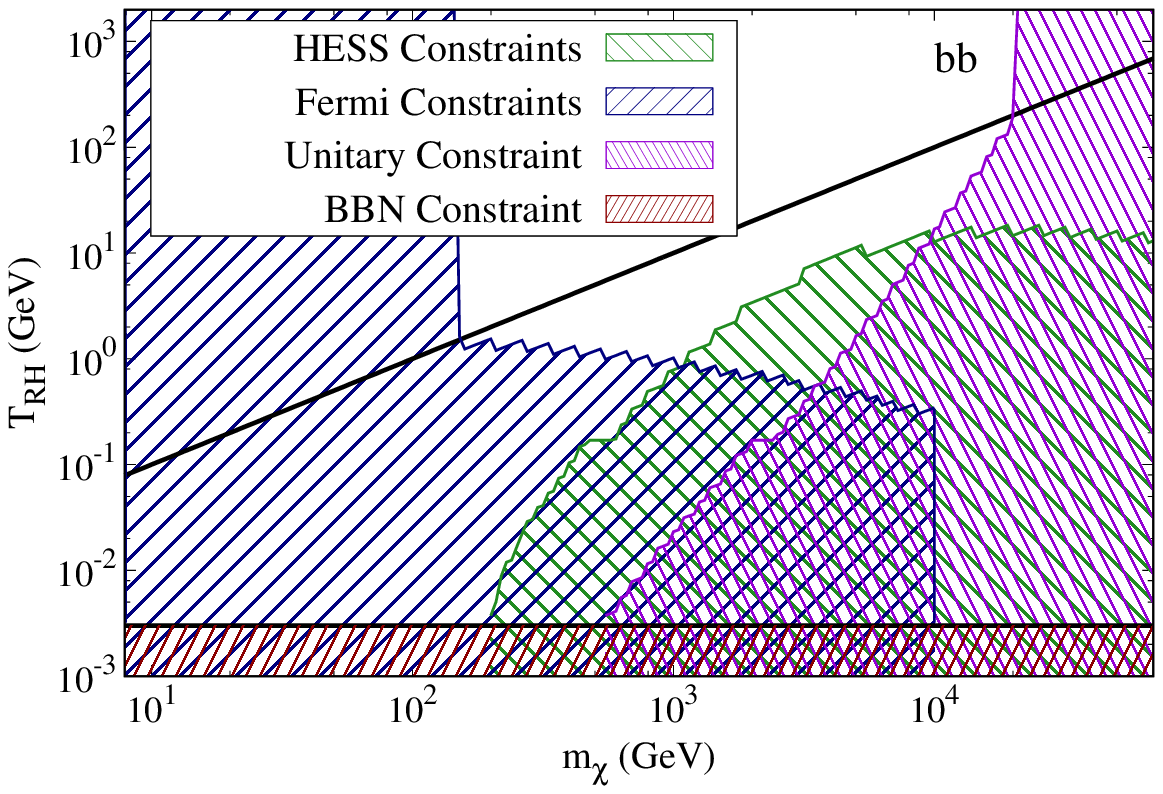}
 \end{subfigure}%
\begin{subfigure}
\centering
 \includegraphics[width=.45\textwidth]{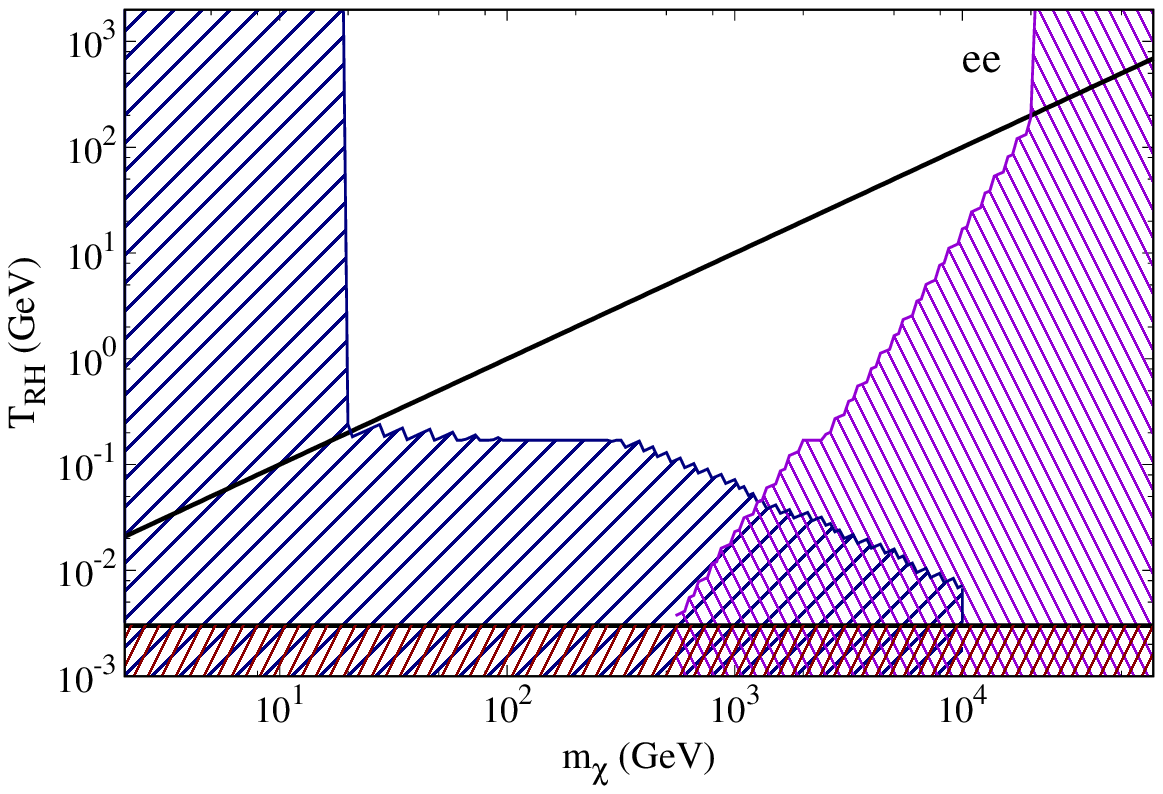}
 \end{subfigure}%
\begin{subfigure}
\centering
 \includegraphics[width=.45\textwidth]{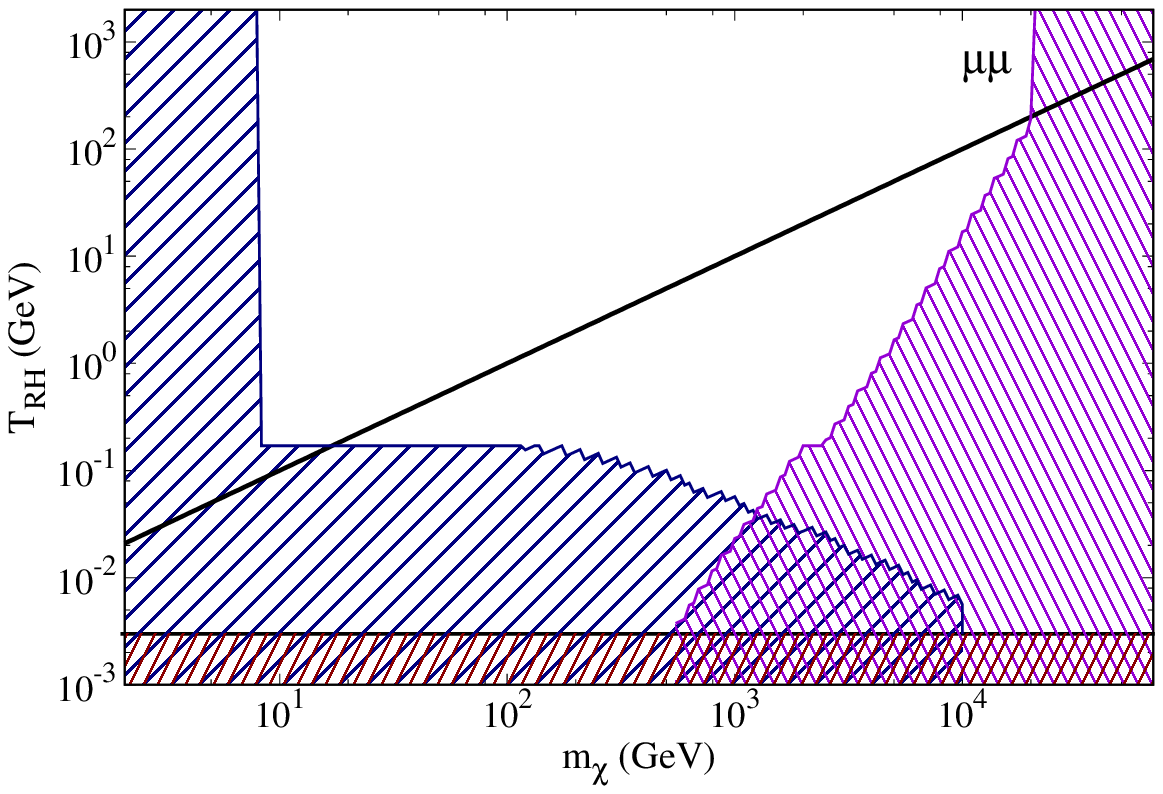}
 \end{subfigure}%
 \begin{subfigure}
\centering
 \includegraphics[width=.45\textwidth]{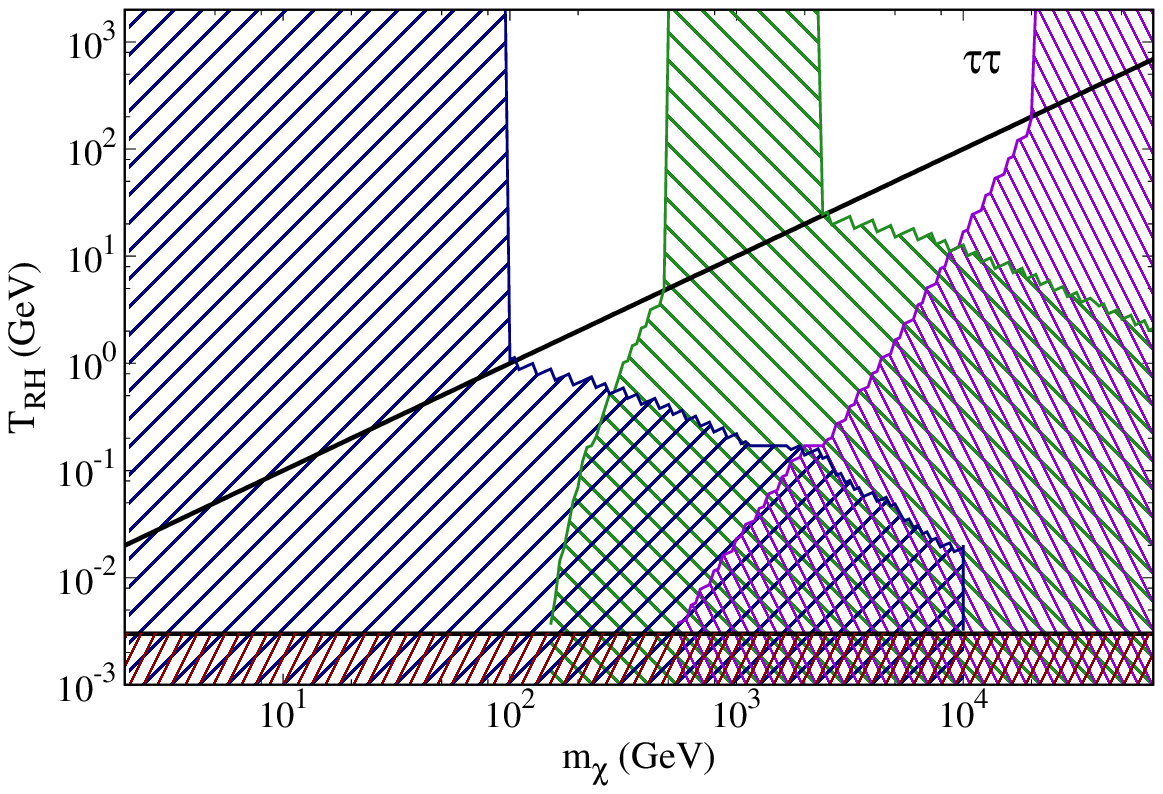}
 \end{subfigure}%
\begin{subfigure}
\centering
 \includegraphics[width=.45\textwidth]{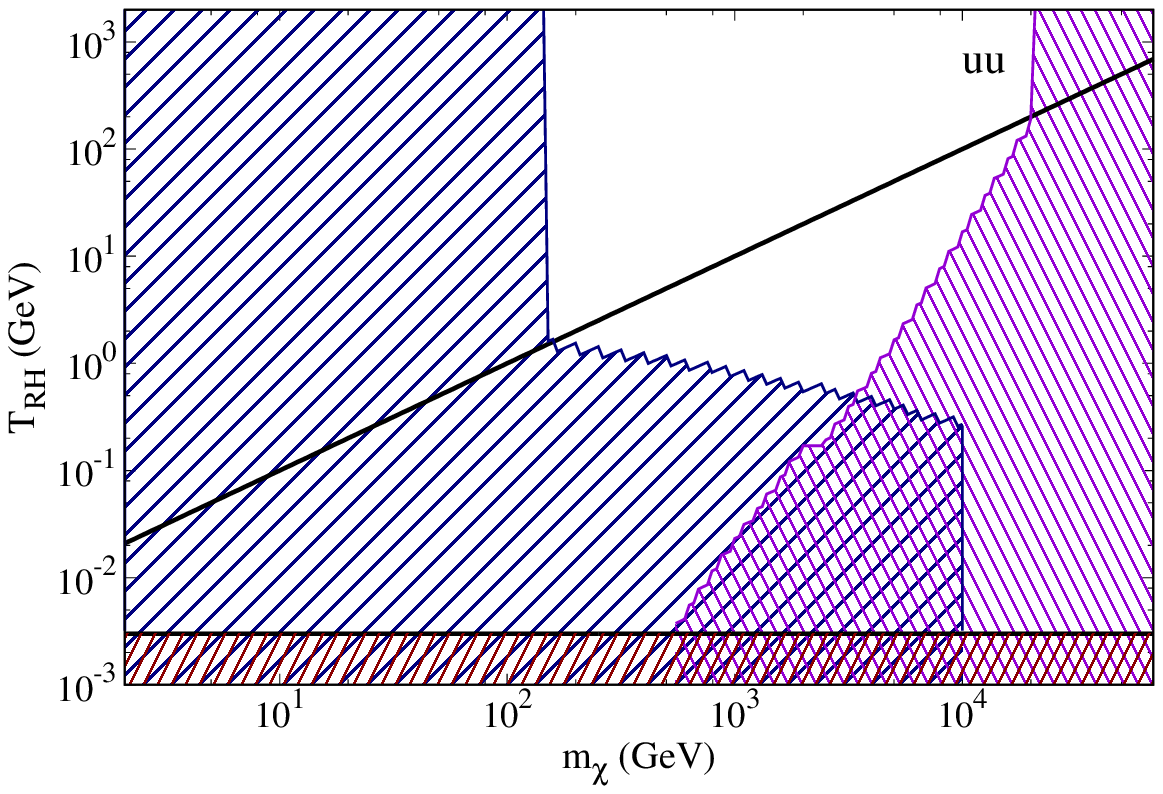}
 \end{subfigure}%
\begin{subfigure}
\centering
 \includegraphics[width=.45\textwidth]{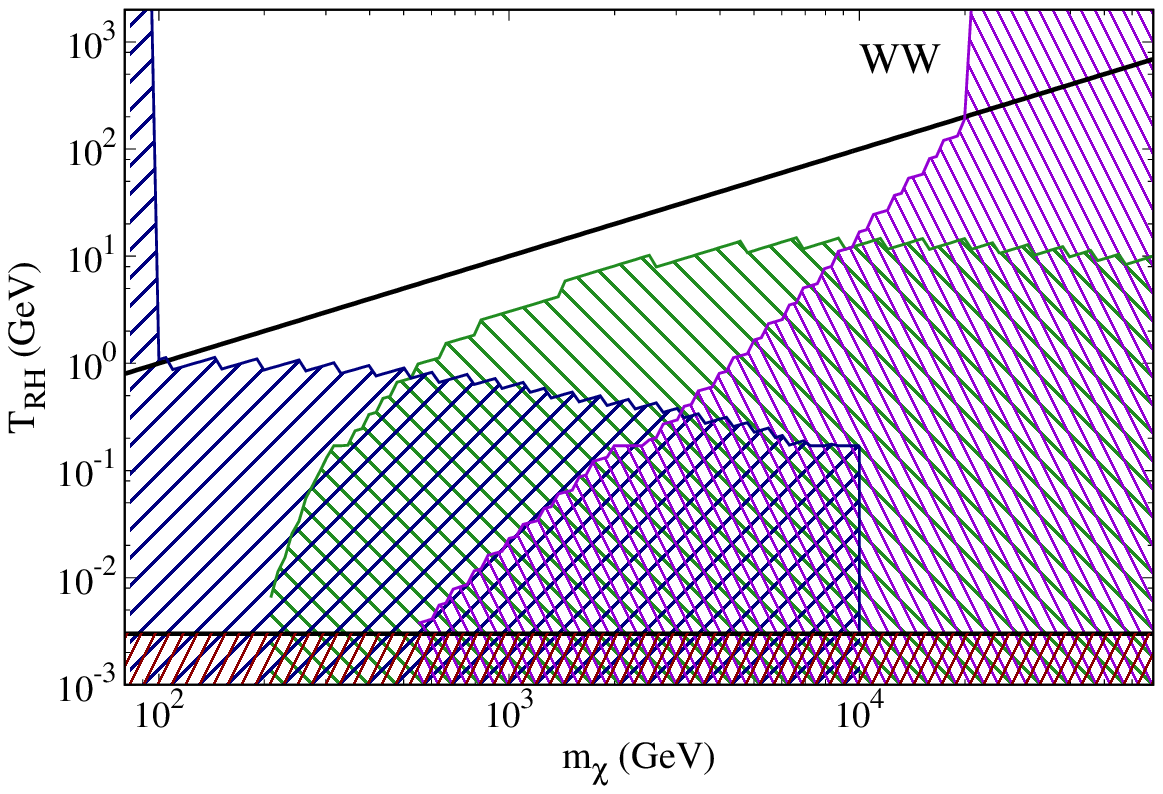}
 \end{subfigure}%

\caption{Constraints on $m_\chi$ and $T_\RH$ for dark matter produced via the freeze-out mechanism.  These panels represent the following annihilation channels:  $b\overline{b}$ (top left), $e^+e^-$ (top right), $\mu^+ \mu^-$ (middle left), $\tau^+ \tau^-$ (middle right), $u\overline{u}$ (bottom left), and $W^+W^-$ (bottom right).  The solid black line represents when ${m_\chi = 100 \, T_{\RH}}$.  Below this line it is possible to reproduce the observed dark matter abundance if dark matter freezes out during kination.  The shaded regions are those that are constrained by Fermi-LAT, H.E.S.S., unitarity, and BBN.}
\label{Fig:KinationConstraints}
\end{figure*}

Figure \ref{Fig:KinationConstraints} shows the Fermi-LAT, H.E.S.S., and unitarity constraints on $m_\chi$ and $T_\RH$ for scenarios in which dark matter freezes out during kination for various annihilation channels.  For every value of $m_\chi$ and $T_{\RH}$ we calculate the $\langle \sigma v \rangle$ that will produce the observed dark matter abundance via freeze-out using Eq.~(\ref{eq:FreezeoutRelicAbundance}).  If the calculated $\langle \sigma v \rangle$ is above the Fermi-LAT or H.E.S.S. constraints, then that scenario is ruled out.  The ruled-out area below ${T_{\RH} = 3 \, \mathrm{MeV}}$ represents the fact that, in order to produce the correct abundance of light elements, reheating must occur before a temperature of ${\sim 3 \, \mathrm{MeV}}$.  The solid black line represents when ${m_\chi = 100 \, T_{\RH}}$.  As discussed in Section \ref{sec:FreezeOut}, Eq.~(\ref{eq:FreezeoutRelicAbundance}) is accurate for ${T_\RH < m_\chi/100}$.  As $T_\RH$ increases beyond $m_\chi/100$, numerical tests with ${m_\chi > 17 \, \mathrm{GeV}}$ indicate that the $\langle \sigma v \rangle$ value that yields the observed dark matter abundance rapidly decreases to $3\times 10^{-26} \sigunits$ as freeze-out occurs closer to radiation domination.  Therefore, we make the conservative assumption that ${\langle \sigma v \rangle = 3\times 10^{-26} \sigunits}$ will give the observed dark matter abundance if ${m_\chi > 17 \, \mathrm{GeV}}$ and ${T_\RH > m_\chi/100}$.

For ${m_\chi < 17 \, \mathrm{GeV}}$, numerical tests show that Eq.~(\ref{eq:FreezeoutRelicAbundance}) remains accurate for reheat temperatures slightly higher than ${m_\chi/100}$ if reheating occurs after the QCD phase transition.  The QCD phase transition causes a sharp decrease in $g_*$ when ${T= 0.17 \, \mathrm{GeV}}$, and since ${T_\mathrm{RH} > 3 \, \mathrm{MeV}}$, ${g_{*s\mathrm{RH}} = g_{*\mathrm{RH}}}$ and ${H \propto g_{*s\mathrm{RH}}^{-1/2}}$ during kination.  Consequently, the Hubble parameter at a given temperature during kination sharply increases as $T_\mathrm{RH}$ goes below ${0.17 \, \mathrm{GeV}}$, which causes freeze-out to occur earlier.  Therefore, Eq.~(\ref{eq:FreezeoutRelicAbundance}) is  applicable for scenarios with $T_\mathrm{RH}$ slightly higher than $m_\chi/100$ if ${m_\chi < 17 \, \mathrm{GeV}}$ because freeze-out still occurs during kination.  For most annihilation channels, scenarios that generate the observed relic abundance with ${m_\chi <17 \, \mathrm{GeV}}$ are ruled out by Fermi-LAT constraints.  The exception is dark matter annihilating via $\mu^+ \mu^-$.  Figure \ref{Fig:KinationConstraints} shows that, for dark matter annihilating via $\mu^+ \mu^-$, Fermi-LAT constraints rule out all scenarios with ${m_\chi \lesssim 8 \, \mathrm{GeV}}$.  In addition, if $m_\chi$ is between ${8 \, \mathrm{GeV}}$ and ${17 \, \mathrm{GeV}}$, Fermi-LAT constraints rule out scenarios with ${T_\RH \lesssim 0.17 \, \mathrm{GeV}}$.  In these scenarios, freeze-out occurs during kination even though $T_\RH$ may be higher than ${m_\chi /100}$.  As $T_\RH$ increases beyond ${0.17 \, \mathrm{GeV}}$, numerical tests with ${8 \, \mathrm{GeV} < m_\chi < 17 \, \mathrm{GeV}}$ indicate that the ${\langle \sigma v \rangle}$ value required to obtain the observed dark matter abundance rapidly decreases to $3\times 10^{-26} \sigunits$ as freeze-out occurs closer to radiation domination.

The resulting constraints on $m_\chi$ and $T_\RH$ are contingent on dark matter reaching thermal equilibrium during kination.  Equation~(\ref{eq:CrossSectionBeginningFreezeOut}) indicates that decreasing $T_\RH$ and increasing $g_{*s}(m_\chi/1.5)$ increases the minimum value of $\langle \sigma v \rangle$ that results in dark matter reaching thermal equilibrium.  Therefore, solving Eq.~(\ref{eq:CrossSectionBeginningFreezeOut}) with the minimum reheat temperature of $3 \, \mathrm{MeV}$ and ${g_{*s}(m_\chi/1.5) = 100}$ shows that, if ${\langle \sigma v \rangle > 3\times 10^{-31} \sigunits}$, dark matter will freeze out during kination regardless of $T_\RH$ or $m_\chi$.

The Fermi-LAT and unitarity constraints establish an allowed mass range for each annihilation channel.  The unitarity bound on $\langle \sigma v \rangle$ places an upper bound on the allowed dark matter mass of ${1.9 \times 10^4 \, \mathrm{GeV}}$ for all annihilation channels.  The lower bound on the dark matter mass comes from the Fermi-LAT observations and is between ${8 \, \mathrm{GeV}}$ and ${160 \, \mathrm{GeV}}$, depending on the annihilation channel.  As $T_\RH$ decreases, the range of viable masses decreases.  The addition of the H.E.S.S. constraints restrict dark matter annihilating via $\tau^+ \tau^-$ to have a mass around either $250 \, \mathrm{GeV}$ or $9000 \, \mathrm{GeV}$.  For dark matter masses between $470 \, \mathrm{GeV}$ and $2500 \, \mathrm{GeV}$, the H.E.S.S. observations constrain $\langle \sigma v \rangle$ to be less than $3\times 10^{-26} \sigunits$ for dark matter annihilating via $\tau^+ \tau^-$, thereby ruling out all scenarios where dark matter freezes out during kination or radiation domination.

Figure \ref{Fig:KinationConstraints} also shows that with the Fermi-LAT, H.E.S.S., and unitarity constraints we can place lower limits on $T_\RH$ if dark matter reaches thermal equilibrium during kination.  For example, we can rule out kination scenarios with reheat temperatures below ${0.05 \, \mathrm{GeV}}$ for dark matter annihilating via the $e^+e^-$ or $\mu^+ \mu^-$ annihilation channel.  We are also able to rule out kination scenarios with reheat temperatures below ${0.6 \, \mathrm{GeV}}$ for the $\tau^+ \tau^-$ and $u\overline{u}$ annihilation channels as well as reheat temperatures below ${1 \, \mathrm{GeV}}$ for the $b\overline{b}$ and $W^+W^-$ annihilation channels.  In addition, kination scenarios where dark matter annihilates via the $b\overline{b}$, $\tau^+ \tau^-$, or $W^+W^-$ annihilation channel require $T_\RH$ be very close to $T_\F$, which implies that these kination scenarios are on the verge of being ruled out.

Throughout this work, we assumed that dark matter consisted of one particle species.  If dark matter consists of multiple particle species, then it is possible that only a fraction of the dark matter is thermally produced during kination.  To determine what effect this has on the $T_\mathrm{RH}$ constraints shown in Figure \ref{Fig:KinationConstraints}, we neglect the ln${(T_\mathrm{F}/T_\RH)}$ term in Eq.~(\ref{eq:FreezeoutRelicAbundance}) and make the rough estimate that for dark matter freezing out during kination, ${\Omega_\chi h^2 \propto m_\chi/\left(\langle \sigma v \rangle \, T_\mathrm{RH}\right)}$.  If only a fraction of the dark matter consists of a thermal relic, such that ${\Omega_\chi h^2 = f \,\Omega_{dm} h^2}$, then to scale the relic abundance by a factor of $f$ for a fixed dark matter mass requires scaling the product of $\langle \sigma v \rangle$ and $T_\mathrm{RH}$ by a factor of $1/f$.  In addition, since the Fermi-LAT and H.E.S.S. constraints are obtained using the dark matter annihilation rate $\Gamma = \langle \sigma v \rangle \rho_{\chi}^2 / m_\chi^2$, altering $\rho_{\chi}$ will subsequently reduce the annihilation rate by a factor of $f^2$ and raise the maximum allowed annihilation cross section $\langle \sigma v \rangle_\mathrm{max}$ by a factor of $1/f^2$.  Therefore, the minimum allowed reheat temperature ${T_\mathrm{RH,min} \propto \langle \sigma v \rangle T_\mathrm{RH} / \langle \sigma v \rangle_{\mathrm{max}} \propto f^{-1}/f^{-2}\propto f}$.  For example, for dark matter annihilating via $W^+W^-$, the minimum allowed reheat temperature is ${1 \, \mathrm{GeV}}$ if dark matter consists of a single particle species.  If only a fraction of dark matter is thermally produced during kination and $\Omega_{\chi}h^2 = 0.05$, then $f = 0.42$ and the new minimum allowed reheat temperature is roughly $0.34 \, \mathrm{GeV}$.

\section{Conclusion}
\label{sec:end}

Our uncertainty regarding the expansion history of the Universe between the end of inflation and the beginning of BBN allows for the possibility that within this period there was an intermittent era of kination.  In this paper we have investigated the effects that a period of kination has on the thermal production of dark matter.  Previous studies on this topic have required the use of specific kinaton potentials \cite{Pallis:2005, Pallis:2nd2005, Pallis:2006, Lola:2009, Pallis:2009}. Our analysis is independent of the kinaton potential, and therefore our constraints on $m_\chi$, $T_\RH$, and $\langle \sigma v \rangle$ are applicable to all kination models assuming that dark matter consists of one particle species that undergoes \textit{s}-wave annihilation.  In addition to numerically solving for the dark matter relic abundance, we have also derived analytic relic abundance equations for freeze-out and freeze-in kination scenarios.

Our scenarios are determined by three parameters: the dark matter mass $m_\chi$, the reheat temperature $T_\RH$, and the dark matter annihilation cross section $\langle \sigma v \rangle$.  In deriving relic abundance equations for freeze-out (Eq.~\ref{eq:FreezeoutRelicAbundance}) and freeze-in (Eq.~\ref{eq:FreezeInRelicAbundance}) scenarios, we have deduced physical relationships between our parameters and the dark matter relic abundance.  For example, at a given temperature, the Hubble parameter during kination is higher than that during radiation domination.  Therefore, freeze-out occurs earlier during kination, which increases the relic abundance.  In order to compensate for this larger relic abundance, freeze-out scenarios require larger-than-canonical $\langle \sigma v \rangle$ values in order to increase the annihilation rate and subsequently decrease the dark matter abundance to the observed value.  If on the other hand, dark matter freezes in during kination, an increase in $T_{\RH}$ will increase the relic abundance.  Increasing $T_{\RH}$ decreases $\rho_{\phi}$, which decreases the Hubble parameter.  Decreasing the Hubble parameter decreases the cooling rate, leaving more time for pair production and thereby increasing the relic abundance.  To compensate for the increased cooling rate and bring the dark matter abundance into agreement with the observed value, freeze-in scenarios during kination require larger $\langle \sigma v \rangle$ values compared to during radiation domination.  Overall, to reach the observed dark matter abundance, freeze-out scenarios during kination require $\langle \sigma v \rangle$ values that would underproduce dark matter during radiation domination, whereas freeze-in scenarios require $\langle \sigma v \rangle$ values that would overproduce dark matter during radiation domination.  Therefore, the possibility that dark matter was thermally produced during kination significantly widens the field of potential dark matter candidates.  In particular, thermally produced Winos and Higgsinos, which generally have $\langle \sigma v \rangle > 3\times10^{-26} \sigunits$, could constitute all the dark matter if they freeze out during a period of kination.

Our analytic relic abundance equations allow us to efficiently determine the dark matter parameter space that would result in the observed dark matter abundance.  To ensure that freeze-out occurs before reheating and that reheating occurs at a temperature above $3 \, \mathrm{MeV}$, $\langle \sigma v \rangle$ values between ${2.7 \times 10^{-38} \sigunits}$ and ${3\times 10^{-26} \sigunits}$ are forbidden for all dark matter masses.  Using the PASS 8 Fermi-LAT observations of dwarf spheroidal galaxies, H.E.S.S. observations of the Galactic Center, and the unitarity bound on $\langle \sigma v \rangle$ we further constrain kination models.  The observational bounds and subsequent constraints only apply to freeze-out scenarios because the required $\langle \sigma v \rangle$ values for freeze-in scenarios are below observational thresholds.  From the unitarity constraint, we were able to rule out all kination scenarios with $\langle \sigma v \rangle$ greater than $4.5\times10^{-23} \sigunits$.  These constraints also allowed us to rule out kination scenarios with reheat temperatures below ${0.05 \, \mathrm{GeV}}$ for dark matter annihilating via the $e^+e^-$ or $\mu^+ \mu^-$ annihilation channel.  Similarly, we ruled out kination scenarios with reheat temperatures below ${0.6 \, \mathrm{GeV}}$ for the $\tau^+ \tau^-$ and $u\overline{u}$ annihilation channels as well as reheat temperatures below ${1 \, \mathrm{GeV}}$ for the $b\overline{b}$ and $W^+W^-$ annihilation channels.  Since these new bounds on $T_\RH$ are below the electroweak phase transition, kination could facilitate baryogenesis \cite{Joyce:1996}.  These bounds on $T_\RH$ are contingent on dark matter freezing out during kination.  We have shown that, if ${\langle \sigma v \rangle > 3\times 10^{-31} \sigunits}$, dark matter will freeze out during kination regardless of $T_\RH$ or $m_\chi$.  We also note that we only consider \textit{s}-wave dark matter annihilation.  If we consider a \textit{p}-wave process, the annihilation rate in the galaxy would be suppressed relative to the annihilation rate at freeze-out and our bounds would no longer apply.

We have shown that scenarios in which dark matter is thermally produced during kination are not ruled out by current observational limits on the dark matter annihilation cross section. In these scenarios, ${\langle \sigma v \rangle > 3\times 10^{-26} \sigunits}$ is required to generate the observed dark matter abundance via a freeze-out process.  The observed dark matter abundance can be generated by a freeze-in process if ${\langle \sigma v \rangle < 2.7\times 10^{-38} \sigunits}$.  Therefore, our uncertainty regarding the pre-BBN expansion history prevents us from knowing the dark matter annihilation cross section that yields the current dark matter abundance; there exists a degeneracy between the allowed values of $\langle \sigma v \rangle$ and $T_\RH$ that cannot be eliminated by relic abundance calculations alone.  One possible approach to breaking this degeneracy involves studying the evolution of the dark matter perturbations during kination.  The evolution of perturbations during kination will impact the small-scale matter power spectrum.  Therefore, studying how the small-scale matter power spectrum and small-scale structure formation are affected by kination may provide a means to further constrain kination scenarios and reduce the range of viable annihilation cross sections.

\textit{Note added}: While we were finishing this paper, Ref.~\cite{DEramo:2017} appeared on the arXiv.  This paper also considers dark matter production during kination, and it similarly identifies the logarithmic decrease of the dark matter number density between freeze-out and reheating.

\section*{Acknowledgments}
We thank Carisa Miller for her comments on our manuscript.  This work was supported by NSF Grant No. PHY-1417446.

\end{document}